\documentstyle[12pt]{article}
\begin{document}
\baselineskip=22pt plus 0.2pt minus 0.2pt
\lineskip=22pt plus 0.2pt minus 0.2pt
\begin{center}
 \Large
Classical and quantum geometrodynamics of \\
2d vacuum dilatonic black holes\\

\vspace*{0.35in}

\large
Madhavan Varadarajan
\vspace*{0.25in}

\normalsize

Department of Physics,\\
University of Utah,\\
Salt Lake City, UT 84112,\\
U.S.A.\\

\vspace{.5in}
July 20, 1995\\
\vspace{.5in}
ABSTRACT
We perform a canonical analysis of the system of 2d vacuum dilatonic black
holes. Our basic variables are closely tied to the spacetime geometry and
we do not make the field redefinitions which have been made by other
authors. We present a careful discssion of asymptotics in this canonical
formalism. Canonical transformations are made to variables which (on shell)
have a clear spacetime significance. We are able to deduce
the location of the
horizon on the spatial slice (on shell)
from the vanishing of  a combination of canonical data.
The constraints dramatically simplify in terms of the new canonical variables
and quantization is easy. The physical interpretation of the variable
conjugate
to the ADM mass is clarified. This work closely parallels that done by
Kucha{\v r} for the vacuum Schwarzschild black holes and is a starting point
for a similar analysis, now in progress, for the case of a
massless scalar field
conformally coupled to a 2d dilatonic black hole.

\end{center}

\pagebreak

\setcounter{page}{1}

\section{Introduction}

In this paper we clarify aspects of the canonical description
of vacuum 1+1 dimensional dilatonic black holes.
Our analysis closely mirrors that of \cite{karel}, in which vacuum
Schwarzschild  black holes were studied.
The motivation for this work stems from our interest in the quantization
of 4-d systems corresponding to spherically symmetric collapse.
More specifically, we would like to study quantum aspects
of spherically symmetric collapse
of a massless scalar field in general relativity in 4 spacetime dimensions.
This is a difficult task because the classical field equations are not
exactly solvable even though the system is effectively 2 dimensional.
However, the CGHS \cite{cghs} model of 2d dilaton gravity with conformally
coupled scalar fields {\em is} classically exactly solvable and we  hope that
the model can be quantized nonperturbatively. Since the 4-d system of interest
is effectively 2-d we hope to gain insights into nonperturbative
quantization of the 4-d
case from a study of the nonperturbative quantization of the CGHS model
(Note that most quantization efforts for the case in which matter is present,
with a few exceptions such as the work of
\cite{mikovic,jackiw}, have been perturbative in character).

The first step in such a study is to present a clear analysis of
the classical and quantum theory of
vacuum dilatonic black holes. Bearing in mind our motivations, we would
like to cast the analysis in a framework which emphasizes the similarities
of this system with the vacuum Schwarzschild black holes.
Both the classical and quantum theory of
the latter has been clearly analysed in \cite{karel} in a canonical framework.
In this paper we show that the 2d vacuum dilatonic black holes
can be handled using exactly the same approach, which worked
for the Schwarzschild case in \cite{karel}.

In this work, we perform a canonical transformation to new canonical pairs
of variables. One of these pairs consist of the mass of the spacetime and
the spatial rate of change of the Killing time. An additional canonical
transformation results in the Killing time itself, being a canonical variable.
The vanishing of the
constraints are shown to be equivalent to, modulo some subtelities,
the vanishing of two of the new canonical momenta. The true degrees of
freedom are parametrised by a canonical pair, one of which is the mass
of the spacetime. It's conjugate variable has a clear spacetime
interpretation. Quantization of this description is almost trivial because
of the simplified form of the constraints. In particular, in contrast to
 \cite{gegenberg}, we can find a Hilbert space representation
(with an appropriate measure) for the physical operators of the theory.

This work, as we see it, has the following merits.
First, it clarifies
the physical interpretation of the observables of the theory (note that
the observables were also given a physical interpretation in \cite{gegenberg};
we go a little further than \cite{gegenberg} in that we discuss parametrisation
of the times at infinity). Second, the initial choice of
variables used in this work is
closely related to the
spacetime geometry of the black hole and the whole treatment possesses
a very close similarity to that in \cite{karel} which deals with the vacuum
Schwarzschild case.
Third,
it can be viewed as a  prerequisite for a similar treatment of the
more interesting conformally coupled matter case.

Along the way, we give a
 self consistent treatment of  asymptotics in the canonical framework (note
the analysis in \cite{asymptotics} is inconsistent due to a subtle technical
reason - this will be pointed out in Section 4).

The layout of the paper is as follows. In section 2, we review the global
structure and the spacetime solution in the
conformal gauge, of the vacuum dilatonic black holes.
In section 3,
we perform the canonical analysis of the action, in analogy to the
ADM analysis done in 3+1 dimensions. Our canonical variables are closely
tied to the  geometry of the black hole spacetime and we {\em do not}
make the field redefinitions of \cite{mikovic}. In section 4, we give a
careful
treatment of asymptotics in the canonical framework and identify the total
energy of the system
with the generator of time translations at spatial infinity. In the rest of
the paper we closely mimic the treatment in \cite{karel}. The idea in
\cite{karel} was to use as canonical variables, quantities which were
physically significant. One hoped that the constraints of the theory
simplified when written in terms of these quantities. The latter were
identified by comparing the spacetime line element written in geometrically
preferred coordinates with the ADM line element.
Thus, in section 5, we express the mass and spatial rate of change of Killing
time in terms of canonical data.  In section 6  we
use these quantities as canonical variables and see
that the constraints simplify when written in terms of these variables.
In section 7,
we bring the `times at infinity' into the canonical framework
exactly as  in \cite{karel}. Finally, in section 8, we give a brief
description of the quantum theory based on the classical decsription of
section7. In section 9 we describe the classical theory by using
light cone coordinates as canonical
variables and quantize this description. Section 10 contains conclusions.

We have not attempted to review the enormous amount of pertinent literature
and we refer the reader to review articles such as \cite{strominger}.

\section{The spacetime solution}
Disregarding boundary terms, the action we deal with is that of
\cite{strominger}:
\begin{equation}
S_{D}\;=\; {1\over{4}}\int d^2x\sqrt{-g}(e^{-2\phi}[{\bf R}
                             + 4(\nabla \phi)^2 + 4K^2])
\end{equation}
where ${\bf R}$
is the scalar curvature of the 2 metric $g_{ab}$, $\phi$ is the
dilaton field and $K$ is the cosmological constant. In this paper we use units
in which $c={2K^2G\over \pi} = {\hbar}=1$
where $G$ is the gravitational constant
dimensions,$c$ is the speed of light and $2 \pi\hbar$ is Planck's constant.
With this choice, mass has units of inverse length.

The solution to the field equations in the conformal gauge
is \cite{strominger} \footnote {Our parameter `$M$' is half the parameter `$M$'
 which appears in \cite{strominger}.}
\begin{eqnarray}
ds^2 & = & - \exp{(+2\psi)}\; dU dV   \\
\exp{(-2 \psi)} & = & \exp{(-2 \phi)}\;\;=\;
{2M\over K}\;-\;K^{2} UV  \\
\Rightarrow {\bf R} & = & {{8MK}\over {{2M\over K}-K^{2}UV}}
\end{eqnarray}
The ranges of $(U,V)$ are such that ${2M\over K}-K^{2} UV \geq 0 $,
with the curvature singularity along the curves ${2M\over K}-K^{2} UV=0 $.

Note that $(U,V)$ here are like the null
Kruskal coordinates $(U_s,V_s)$ for
Schwarzschild. In the latter, the curvature
 singularity is at $U_s V_s =1$. Here $K$  provides an extra scale, which
allows for the definition of dimensional Kruskal like coordinates,
in contrast with the dimensionless Kruskal coordinates for Schwarzschild.

The global structure of the vacuum dilatonic black hole is identical to that
of the fully extended Schwarzschild solution. We label the different parts
of the spacetime as follows: Region I with $U<0, V>0$
(the right static region), Region II with $U>0, V>0$ (the future dynamical
region), Region III with $U>0,V<0$ (the left static region) and Region IV
with $U<0,V<0$ ( the past dynamical region). The horizons are at $U=0$ and
$V=0$. We can define  $(T,\rho )$  coordinates (the analog of the
Killing time and Regge-Wheeler tortoise coordinates for Schwarzschild)
in regions I and III as follows:
\begin{eqnarray}
KV=e^{K(T+\rho )}\;\; & KU=  -e^{-K(T-\rho )}
    &{\rm in\; region \;I} \\
KV=-e^{K(T+\rho )}\;\; & KU=  e^{-K(T-\rho )}
       &{\rm in \;region\; III}
\end{eqnarray}
In each  of the regions I and III,
the line element and the coordinate ranges are
\begin{equation}
ds^2 =
{{e^{(2K\rho )} }\over {{2M\over K} +e^{(2K\rho)} }}
 [-(dT)^2 +(d \rho )^2 ]    \;\;\;\;- \infty < T,\rho < \infty
\end{equation}
and the horizons are at $\rho \rightarrow -\infty $
with spatial infinity at $\rho \rightarrow \infty $.
The metric is manifestly asymptotically flat at left and right
spatial infinities in these coordinates.

\section{The canonical form of the hypersurface action}

We apply the standard ADM formalism  of general relativity in 3+1 dimensions
to the 2-d dilatonic black hole spacetimes. The spacetime is foliated by
a 1 parameter family of slices `$\Sigma $', the slices being
labelled by  a time parameter `$t$', $t\epsilon R$.
Each $t=$constant slice, $\Sigma$, has
the topology of $R$ and is coordinatized by a parameter `$r$'. Further,
each slice is spacelike and extends from left spatial infinity to right
spatial
infinity, but is otherwise arbitrary (in particular, the slices {\em are not}
restricted to pass through the bifurcation point $(U,V)=(0,0)$ ).\\
\noindent The range of $r$ is $-\infty <r < \infty$ with
left and right spatial infinities being approached
as $r \rightarrow -\infty$ and $r \rightarrow \infty$ respectively.
The ADM form of the spacetime line element is
\begin{equation}
ds^2 \;= \; -(N^2 - (N^{r})^2 \Lambda^{2})(dt)^{2}
               \;+\; 2 \Lambda^{2}N^{r}(dt)(dr) \;+\; \Lambda^{2}(dr)^2 .
\end{equation}
Here, $\Lambda^2 (dr)^{2}$ is the induced spatial metric on $\Sigma$ and
$N$ and $N^{r}$ are the usual lapse and shift parameters.
We substitute this form of the spacetime metric into the action $S_D$
and obtain, modulo boundary terms
\begin{eqnarray}
S_{\Sigma} & = & \int dt dr({-1\over N}
   [(-{\dot {\Lambda}}+(N^{r}\Lambda )^{\prime})
    (-{\dot R}+N^{r}R^{\prime})R
\;\;+\;\Lambda (-{\dot R}+N^{r}R^{\prime})^2]  \nonumber \\
      & + & N[-{ RR^{\prime \prime}\over \Lambda }
              \;+ \;{ \Lambda^{\prime} R R^{\prime} \over \Lambda^2}
              \;+\; K^2 R^2 \Lambda])
\end{eqnarray}
where dots and primes
 denote derivatives with respect to $t$ and $r$ respectively and
we have defined $R:= e^{-\phi}$ to facilitate comparision with
\cite{karel}. We shall supplement the above hypersurface action with
the appropriate boundary terms in section 4.

The next step in the canonical analysis is to identify the momenta
conjugate to the variables $R, \Lambda$ ($N, N^a$ will turn out to be
Lagrange multipliers). The momenta are
\begin{eqnarray}
P_{\Lambda} & = \; {\delta S_D \over \delta {\dot \Lambda}} &
              = \;{1 \over N}
(-{\dot R}+N^{r}R^{\prime})R \\
P_{R} & = \; {\delta S_D \over \delta {\dot R}} &
              =\; {1 \over N}[(-{\dot {\Lambda}}+(N^{r}\Lambda )^{\prime})R
   \; +\;2\Lambda (-{\dot R}+N^{r}R^{\prime})]
\end{eqnarray}
The canonical form of the hypersurface action is
\begin{equation}
S_{\Sigma}[\Lambda,P_{\Lambda},R,P_{R};N,N^{r}]\;=\;
     \int dt \int_{-\infty}^{\infty}dr
(P_{\Lambda} {\dot \Lambda}\;+\;P_R {\dot R}\;-\;NH\;-\;N^{r}H_{r})
\end{equation}
where
\begin{eqnarray}
H &= & {P_{\Lambda}^{2} \Lambda \over R^2}\; -\; {P_{R}P_{\Lambda} \over R}
       \; +\;{ RR^{\prime \prime}\over \Lambda }
              \;- \;{ \Lambda^{\prime} R R^{\prime} \over \Lambda^2}
              \;-\; K^2 R^2 \Lambda   \\
H_{r} & = & P_{R}R^{\prime}- P_{\Lambda}^{\prime}\Lambda
\end{eqnarray}
$H$ and $H_r$ are the Hamiltonian and diffeomorphism constraints
of the theory. $N,N^r$ are Lagrangian multipiers.
The symplectic structure  can be read off from the action and
the only non vanishing Poisson brackets are
\begin{equation}
 \{\Lambda (y) , P_{\Lambda}(x)\} =
\{R (x) , P_{R}(y)\} = \delta (x,y)
\end{equation}
It is straightforward to check that with these Poisson brackets the
constraints are first class. It is also easy to see that  $H_r$ integrated
against $N^r$
generates spatial diffeomorphisms
of the canonical data.

\section{Asymptotics}

A satisfactory treatment of asymptotics in the context of a Hamiltonian
formalism must achieve the following:\\
\noindent (i) The choice of asymptotic conditions on the
fields must be such that canonical data corresponding to classical solutions
of interest are admitted.\\
\noindent(ii)Under evolution or under flows generated by the constraints
of the theory, the canonical data should remain in the phase space.
For example, the smoothness of the functional derivatives of the generating
functions for such flows must be of the same type as the data themselves.
Also, the boundary conditions on the data must be preserved by such flows.
The latter requirement may be violated in subtle ways and the formalism
must be carefully checked to see that this does not happen.

In what follows we display our choice of boundary conditions on the phase
space variables. To impose (ii), above, we need to deal with appropriate
functionals on the phase space.  These are obtained by integrating the local
expressions  for $H$ and $H_r$ against the multipliers $N$ and $N^r$ to
obtain
$C(N):=\int_{-\infty}^{\infty}dr NH$
 and $C(N^r):=\int_{-\infty}^{\infty}dr N^rH_r$. Imposition of (ii)
gives  boundary conditions on the $N$ and $N^r$. Next, we identify
the generator of unit time translations at spatial infinity and show that
it is  $M$ on shell( $M$ is the parameter in the spacetime metric).
Finally, we point out why  the analysis in \cite{asymptotics}
is incomplete.

Since we use the same techniques as in the analysis for 3+1 canonical gravity,
we shall not discuss them in detail.

\subsection{Boundary conditions on phase space variables}
At right spatial infinity $(r\rightarrow \infty)$ we impose:
\begin{eqnarray}
R & = & e^{Kr}\; + \; \alpha_+ e^{-Kr} \; + \;O(e^{-3Kr}) \\
\; & \Rightarrow & R^{2}=e^{2Kr}[1\;+\;2\alpha_+ e^{-2Kr} +\;O(e^{-4Kr})] \\
\Lambda & = & 1 - \; \beta_+ e^{-2Kr} \; + \;O(e^{-4Kr}) \\
   \;   & \Rightarrow & \Lambda^{2}= 1\;-\;2\beta_+ e^{-2Kr} +\;O(e^{-4Kr})
\end{eqnarray}
\begin{eqnarray}
P_{R} & \sim & e^{-Kr} \\
P_{\Lambda} & \sim & O(1)
\end{eqnarray}
Here $\alpha_+$ and $\beta_+$
are arbitary (in general time dependent) paramters
independent of $r$ .

At left spatial infinity ($r \rightarrow -\infty$) we impose exactly
the same conditions except that $r$ is replaced by $-r$ and the `right'
parameters
$\alpha_+$ and $\beta_+$ are replaced by
the `left' parameters denoted by $\alpha_-$ and $\beta_-$.
Note that, on solution, $\alpha_+ = \alpha_-
= \beta_+ = \beta_- ={M\over K}$.

\subsection{The diffeomorphism constraint functional}

We require that \\
\noindent (a)$C(N^r)$  exists. From the boundary conditions (16)-(21), above
\begin{equation}
H_{r}\sim O(1)
\end{equation}
Thus, for $C(N^r)$ to be finite, the asymptotic behaviour of $N^r$ should be
\begin{equation}
N^r \sim O({1\over r^2})
\end{equation}
\noindent (b)The orbits generated by $C(N^r)$ should lie within the phase
space. Consider
\begin{eqnarray}
{\dot R}(x)& := & \{R(x),C(N^r)\} = N^r(x) R^{\prime}(x) \nonumber\\
\Rightarrow {\dot R} & \sim & N^r e^{K|r|} \;\;,|r|\rightarrow \infty
\end{eqnarray}
But for $R(t,x)$ to satisfy the boundary condition (16), asymptotically
\begin{eqnarray}
{\dot R} & \sim  & O(e^{-K|r|}) \nonumber \\
\Rightarrow N^r & \sim & e^{-2K|r|}
\end{eqnarray}
This is stronger than (23). For $N^r$ satisfying (25),
it can be verified that
$C(N^r)$ generates orbits which preserve the boundary conditions
on the canonical variables and is functionally differentiable without the
addition of a surface term. Since asymptotic translations of $r$
 do not preserve the the boundary conditions on $R$, there are no generators
of such translations in the theory (if there were, we would identify them
with spatial momenta).

\subsection{The Hamiltonian constraint functional and the mass}

We require that $C(N)$ exists. Using the boundary conditions on the
canonical variables, we find that $H\sim e^{-2K|r|}$
near the spatial infinities. For each of
the momentum dependent terms in $H$ this
is obvious. The remaining terms, individually, do not have this behaviour;
but taken together, conspire to fall off as $e^{-2K|r|}$. So, for $C(N)$
to exist, it is possible to admit
\begin{eqnarray}
N & \rightarrow &N_+ ,\;\; \;\; r\rightarrow +\infty \\
N & \rightarrow &N_- ,\;\;\;\; r\rightarrow -\infty
\end{eqnarray}
where $N_+, N_-$ are independent of $r$.

Next, we require that $C(N)$ be functionally differentiable. The variation of
$C(N)$ is of the form:
\begin{equation}
\delta C(N) =
      \int_{-\infty}^{\infty}dr[
   \Xi_\Lambda \delta \Lambda (r) +\Xi_R \delta R(r) +
     \Xi_{P_\Lambda} \delta P_\Lambda (r) +\Xi_{P_R} \delta P_{R}(r) ]
 + \delta C(N)_{\rm surface}
\end{equation}
Here $\Xi_R, \Xi_{P_R}, \Xi_{\Lambda}, \Xi_{P_{\Lambda}}$ are local
expressions
involving the canonical variables and the lapse and their derivatives. The
term which obstructs functional differentiability is
$\delta C(N)_{\rm surface}$ and is given by
\begin{equation}
\delta C(N)_{\rm surface} = s(r=\infty )- s(r=-\infty )
\end{equation}
where
\begin{equation}
s\; =\; -{NRR^{\prime}\over \Lambda^2} \delta \Lambda
    -{(RN)^{\prime}\over \Lambda}\delta R
    -{RN\over \Lambda}(\delta R)^{\prime}
\end{equation}
$\delta C(N)_{\rm surface}$ vanishes if $N\rightarrow 0$ as
$|r|\rightarrow \infty$ and $C(N)$ is functionally differentiable for
such $N$.

To complete the analysis for the Hamiltonian constraint
functional, we must check that the boundary conditions are preserved by the
flows generated by $C(N)$. We have
\begin{eqnarray}
{\dot P}_{\Lambda} & := & \{ P_{\Lambda}, C(N)\}\;=\; -
\Xi_{\Lambda}\nonumber\\
         &= & {NP_{\Lambda}^{2} \over R^2}
       \; +\;{ NR^{\prime 2}\over \Lambda^2 }
              \;+\;{ N^{\prime} R R^{\prime} \over \Lambda^2}
              \;-\; NK^2 R^2
\end{eqnarray}
For the boundary conditions to be satisfied, we must
 require ${\dot P}_{\Lambda} \sim O(1)$
asymptotically. From the  equation for ${\dot P}_{\Lambda}$ and the
boundary conditions on the canonical variables, it can be verified that this
requirement implies
\begin{equation}
N^{\prime} \sim e^{-2K|r|} ,\;\; |r| \rightarrow \infty
\end{equation}
This fixes $N\sim e^{-2K|r|}$ asymptotically. It can be checked that for
such $N$ , $C(N)$ exists, is functionally differentiable and generates
flows which preserve the boundary conditions.

Thus, just as in canonical gravity in 3+1 dimensions, the constraints
$C(N), C(N^r)$ generate motions which are  trivial at  infinity.
We have seen that the spatial momentum vanishes because constant spatial
translations at infinity do not preserve the asymptotic conditions.
We now turn our attention to constant time translations at the spatial
infinities and identify the generators of such motions with the ADM
masses.

Hence, we must impose  (26) and (27). (32) still holds since we want the
boundary conditions on $P_{\Lambda}$ to be preserved under evolution.
Thus
\begin{eqnarray}
N & \rightarrow &N_+  + O(e^{-2Kr}),\;\; r\rightarrow +\infty \\
N & \rightarrow &N_- + O(e^{2Kr}),\;\; r\rightarrow -\infty
\end{eqnarray}
where $N_+, N_-$ are independent of $r$.
$C(N)$ is not functionally differentiable due to the presence of
$\delta C(N)_{\rm surface}$. This term, for $N$ with the above behaviour is
\begin{equation}
\delta C(N)_{\rm surface} =-K(2\delta \alpha_+ -\delta \beta_+)N_+
                    - K(2\delta \alpha_- -\delta \beta_-)N_-
\end{equation}
To restore functional differentiability, we add an appropriate term to
$C(N)$ whose variation cancels $\delta C(N)_{\rm surface} $. We call the
resultant expression, (which is non zero on the constraint surface) the true
Hamiltonian $H_T$. It is given by
\begin{equation}
H_{T}(N):= C(N) + K(2 \alpha_+ -\beta_+)N_+
                    +K(2 \alpha_- - \beta_-)N_-
\end{equation}
$H_T$ generates time translations which are constant
at right and left spatial
infinities and we can identify
\begin{equation}
M_+ := K(2\alpha_+ - \beta_+)
\end{equation}
and
\begin{equation}
M_-  :=  K(2\alpha_- -\beta_-)
\end{equation}
 with the masses
at right and left infinity, respectively.
(Note, that on shell these expressions both reduce
to the parameter $M$ in the spacetime solution.)

Thus, the correct action to use is
\begin{eqnarray}
S[\Lambda, P_\Lambda, R,P_R,N,N^r] & = &
         S[\Lambda, P_\Lambda, R,P_R,N,N^r] \nonumber\\
                        & + & S_{\partial \Sigma}[M_-,M_+,N_-,N_+]
\end{eqnarray}
where
\begin{equation}
S_{\partial \Sigma}[M_-,M_+,N_-,N_+] =  -\int dt (N_+M_+  +N_-M_-)
\end{equation}

\vspace{2mm}

 The analysis in \cite{asymptotics} is incomplete
for the following reason. In \cite{asymptotics} the boundary conditions are
a little stronger than ours - in particular, the work there assumes
$\alpha = \beta$ (we have suppressed the $+$ and $-$ subscripts).
 However the true
Hamiltonian {\em does not} necessarily
preserve $\alpha =\beta$. This can be seen by
looking  at the evolution of $R$ and $\Lambda$:
\begin{equation}
{\dot R} = \{ R,H_T(N)\}= -{NP_{\Lambda}\over R}
\end{equation}
\begin{equation}
{\dot \Lambda} = \{ \Lambda,H_T(N)\}=-{NP_{R}\over R}
                                         +{2NP_{\Lambda}\Lambda\over R^2}
\end{equation}
Our boundary conditions, on the other hand, suffer no such deficiency.

\section{Reconstruction of mass and Killing time rate from canonical data}

We would like to construct, out of the canonical data, expressions which
on shell, give the mass, $M$ and spatial rate of change of Killing time,
$T^{\prime}$, along $\Sigma$. We guess the correct expressions, just as in
\cite{karel}, by comparing the ADM form of the line element (8)
with that corresponding to the spacetime solution (7). So we
parmeterize $T=T(t,r)$ and $R=R(t,r)$ and put them into (8). Comparision
with (7) gives:
\begin{eqnarray}
\Lambda^{2} & = & \Phi^{2}(-T^{\prime 2}\; +\;\rho^{\prime 2}) \\
N^r & = & {-{\dot T}T^{\prime}\;+\;{\dot \rho}\rho^{\prime}
                   \over -T^{\prime 2}\; +\;\rho^{\prime 2}} \\
N & = & \Phi {{\dot T}\rho^{\prime}\;-\;{\dot \rho}T^{\prime}
                \over \sqrt{-T^{\prime 2}\; +\;\rho^{\prime 2}}} \\
{\rm where} & \Phi^2 & :={{e^{(2K\rho )} }\over {{2M\over K} +e^{(2K\rho)} }}
                                 \\
{\rm Note \;that} & R^2 \; = &
 {2M\over K} +e^{(2K\rho)} \;{\rm on\; solution\;}
\end{eqnarray}
(To understand the
reasons for our choice of signs for the square root in the expression for $N$
 see  \cite{karel}).
Substituting this in the expression for $P_{\Lambda}$ (10) we get
an expression for the Killing time rate
\begin{equation}
T^{\prime}\; =\; {-P_{\Lambda} \Lambda \over K(R^2\;-\;{2M\over K})}
\end{equation}
Finally, making use of equation (43) for $\Lambda^2$ , we get an expression
for the mass
\begin{equation}
{2M\over K} \;= \; R^{2} \; +\; {P_{\Lambda}^{2}\over K^2 R^2}\;-\;
            {R^{\prime 2}\over \Lambda^2 K^2}
\end{equation}
Thus we have an expression for the mass as a function of the canonical data
and we can substitute this expression for $M$ in the above expression for
$T^{\prime}$ to get the Killing time rate also as  function of the
canonical data. Note the similarity of the expressions with those in
\cite{karel}.
Finally, from the boundary conditions on
the canonical data (16)-(21), we can infer the asymptotic behaviour
of our  expression for the mass. We obtain
\begin{equation}
M(r=+\infty ) =K(2\alpha_+ -\beta_+)
\;\;\;\;
M(r=-\infty ) =K(2\alpha_- -  \beta_-)
\end{equation}
This is exactly what we expect from the expressions for the generators
of unit time translations at spatial infinities derived in
section 4.3 !

\section{Using $M$ and $T^{\prime}$ as new canonical variables}
\subsection{The canonical transformation}
It is straightforward to show that
\begin{equation}
\{ M(x), -T^{\prime}(y) \} \;=\; \delta (x,y)
\end{equation}
This prompts the definition
\begin{equation}
P_M(x)\; := \;-T^{\prime}(x)
\end{equation}
Since neither $M(x)$ nor $P_M(x)$ contain $P_{R}$, they commute under
Poisson brackets with $R$. However $(M,P_M; R,P_R)$ are not a canonical
chart and we need to replace $P_R$ with an appropriate variable to have
a canonical chart on phase space.  We use the same trick as in \cite{karel}
to guess the new momentum conjugate to $R$ (which we shall refer to as
$\Pi_R$). We expect $\Pi_R$ to be a density of weight one and require the
diffeomorphism constraint to generate diffeomorphisms on the new canonical
variables. This prompts the definition
\begin{equation}
\Pi_{R}\; := \; P_R \; -\; {1\over R^{\prime}}(M^\prime P_M \;+\;
                           \Lambda P_{\Lambda}^{\prime})
\end{equation}
Long and straightforward calculations show that
\begin{equation}
\{ R(x), \Pi_R(y) \}= \delta(x,y)\;\;\;\{\Pi_R(x),M(y)\}
                                       =\{\Pi_R (x),P_{M}(y)\}
               =\{\Pi_R (x),\Pi_{R}(y)\}=0
\end{equation}
Thus, we make a canonical transformation from $(\Lambda, P_\Lambda ;R,P_{R})$
 to $(M,P_{M}; R, \Pi_{R})$. We can equally well express the old variables
in terms of the new.
\begin{eqnarray}
\Lambda^{2} & = & {R^{\prime 2} \over fK^2}\;-\; {P_{M}^{2} f\over R^2} \\
P_{\Lambda} & = & {KfP_M \over ({R^{\prime 2}\over fK^2} \;-\;
                                               {P_{M}^{2}f \over R^{2}})
                                                 ^{1\over 2}}
\end{eqnarray}
Here we have defined
\begin{equation}
f\; := \; R^2 - {2M \over K} ={R^{\prime 2} \over (\Lambda K)^2}
                             - {P_{\Lambda}^{2}\over K^2 R^2}
\end{equation}
{}From the spacetime solution, it is apparent that (on shell), the horizons
are  located at $f=0$. For $f=0$ the canonical transformation breaks down.
This is exactly what happens in the Shwarzschild case and we refer the
reader to \cite{karel} for a discussion of issues which arise when $f=0$.

\subsection{The constraints}
Before writing the constraints in terms of the new canonical
variables, it is instructive to show that $M(x)$ is a constant of motion.
{}From (49) and (13,14) it is easy to check that
\begin{equation}
{M^{\prime} \over K}= -{1\over R\Lambda K^2}(R^{\prime}H\;+\;
                               {P_{\Lambda}\over R}H_{r})
\end{equation}
Thus, as expected, the mass function doesn't change over the slice.
It is also easy to check that
\begin{equation}
\{M(x), H(y) \}\;=\; -\delta (x,y) {R^{\prime} \over \Lambda^3 K R} H_{r}
\end{equation}
as well as
\begin{equation}
\{M(x), H_r(y) \}\;=\; M^{\prime}(x) \delta (x,y)
\end{equation}
(59), (60) and (58) show that $M(x)$ is indeed a constant of motion.

We now proceed to write the constraints in terms of the new canonical
variables. From the expressions for $\Pi_R$ ,(53), and $M^\prime$,(58), it
can be shown that
\begin{eqnarray}
H & = & - { {M^{\prime}RR^{\prime}\over fK} + {KfP_M \Pi_R \over R}
                      \over ({ R^{\prime 2}\over fK^2}-
                                 {P_{M}^{2}f \over R^2})^{1\over 2} }\\
H_{r} & = &  \Pi_R R^{\prime} + P_M M^{\prime}
\end{eqnarray}

Thus, the vanishing of  $H,H_r$ is equivalent to the vanishing of
$M^\prime, \Pi_R$ modulo the vanishing of $f$. Again, arguing the same way as
in \cite{karel}, we can replace, as constraints, the former
with the latter everywhere on $\Sigma$ (in \cite{karel} the argument for
replacing the old constraints with the new even when $f$=0 is essentially one
of continuity). We can express the old set of constraints in terms of the
new ones  (61,62)  in matrix notation i.e.:

\begin{equation}
\left[
\begin{array}{c}
H\\
H_{r}\\
\end{array}
\right] \;=\; A\left[
\begin{array}{c}
M^{\prime}\\
\Pi_{R}\\
\end{array}
\right]
\end{equation}
where $A$ is a $2\times 2$ matrix whose field dependent coefficients
can be read of from (61,62) above.
It is curious that even though the individual elements  of  $A$
maybe ill defined when $f=0$, its determinant is independent of $f$. In fact
\begin{equation}
{\rm Det}A\;=\;RK\Lambda
\end{equation}
and is non vanishing as long as $R, \Lambda \;  \neq 0$. Since $\Lambda^2$
is the spatial metric, $\Lambda$ is non zero and $R \geq 0$
by virtue of it's definition in terms of the dilaton field,
vanishing on shell only at the singularity.
So, if $\Sigma$ is away from the singularity, the behaviour of Det$A$
supports the replacement of the old constraints with the new.

\subsection{The action}
 Written in terms of the new canonical variables, the hypersurface
action is
\begin{equation}
S_{\Sigma}[M,R,P_{M},\Pi_{R};N,N^{r}]\;=\;
     \int dt \int_{-\infty}^{\infty}dr
(P_{M} {\dot M}\;+\;\Pi_R {\dot R}\;-\;NH\;-\;N^{r}H_{r})
\end{equation}
Replacing the old constraints with the new ones gives
\begin{equation}
S_{\Sigma}[M,R,P_{M},\Pi_{R};N,N^{r}]\;=\;
     \int dt \int_{-\infty}^{\infty}dr
(P_{M} {\dot M}\;+\;\Pi_R {\dot R}\;-\;N^M M^{\prime}\;-\;N^{R}\Pi_{R})
\end{equation}
where $N^M$ and $N^R$ are new Lagrange multipliers related to the old ones by
\begin{eqnarray}
N^M & = & - {RR^{\prime}\over fK\Lambda}N + P_M N^r \\
N^R & = &  -{KfP_{M} \over R \Lambda}N + R^{\prime}N^r
\end{eqnarray}
{}From (40) , the boundary action
$S_{\partial \Sigma}$  is given by
\begin{equation}
S_{\partial \Sigma} = -\int dt(N_+M_+ + N_-M_-)
\end{equation}
where (from (37,38) and (50))
\begin{equation}
M_+ = M(r=\infty )\;\;\;\;\;M_-= M(r=-\infty )
\end{equation}
Note that from (67,68)
and the boundary conditions on the canonical data and the
lapse and shift functions, the asymptotic values of the new multipliers are
$N^R \rightarrow 0$ and
\begin{eqnarray}
N^M(r=+\infty)& =: \;N^{M}_{+} &\;=-N_+\\
N^M(r=-\infty)& =: \;N^{M}_{-} &\;=+N_-
\end{eqnarray}
 The total action is just
\begin{equation}
S[M,R,P_M,\Pi_R;N^M,N^R]=S_\Sigma [M,R,P_M,\Pi_R;N^M,N^R]
+S_{\partial \Sigma}[M;N^M]
\end{equation}

\vspace{2mm}

We now a raise a subtle issue
(whose analog has gone unnoticed in \cite{karel}). The old variables
$\Lambda, R, P_\Lambda, P_R$  were taken to be smooth functions of $r$,
subject to the boundary conditions in section 4.  The new variables
(specifically) $P_M$ and $\Pi_R$ are {\em not} necessarily smooth functions
of $r$ for arbitrary values of the old variables. In fact, on shell, on the
horizon, if $\Sigma$ does not pass through the bifurcation point, the
new momenta {\em necessarily} diverge. Thus, if we are to admit
{\em arbitrary} slicings, not just the ones passing through the bifurcation
point, we must allow for divergent values of the new momenta (at least
when $f=R^2-{M\over K}$ vanishes). A similar comment holds for
the new Lagrange multipliers. So, if we use the action in terms of
the new  constraints we cannot assume the data to be smooth functions. This
raises questions about the associated quantum theory. In \cite{karel}
the theory is quantized without worrying about this issue - since we do not
know how to resolve this issue we shall also do the same. As a result,
it may be that we are only allowing slices which pass through the bifurcation
point on shell (although, it may be that since $f=0$ only on a
``set of measure zero'',we can choose to ignore the issue).

\section{The parametrized action}

Since the form of the action (and even the notation, with the exception
of the symbol for the momentum conjugate to $R$) is identical to that in
\cite{karel}, one can simply follow the discussion of sections VII to IX
of that
paper. We will be extremely succint in this section and quickly review the
relevant part of \cite{karel}.

For quantization, one can deal with the action in section 6.3 in which the
lapse functions are prescribed at spatial infinities, or one can parameterize
the proper times at right and left infinities
(denoted by $\tau_+$ and $\tau_-$) and obtain the action
\begin{equation}
S[M,P_M, R, \Pi_R, N^M, N^R,\tau_+  , \tau_- ]
:= S_{\Sigma}[M,P_M, R,\Pi_R , N^M, N^R] +
S_{\partial \Sigma}[M,\tau_+, \tau_-]
\end{equation}
where
\begin{equation}
S_{\partial \Sigma}[M,\tau_+, \tau_-]:= -\int dt (M_+ \dot{\tau_+}-
                                                M_- \dot{\tau_-})
\end{equation}
and $S_{\Sigma}[M,P_M, R,\Pi_R , N^M, N^R] $ is given in section 6.3 .
The {\em free}
variations of the parametrized action
with respect to all its arguments result in equations of motion
equivalent to those obtained from the  action in section 6.3 which
had {\em prescribed} lapses at spatial infinities.

Further analysis of the parametrized action
$S[M,P_M, R, \Pi_R, N^M, N^R,\tau_+  , \tau_- ]$ reveals that the following
definitions
\begin{eqnarray}
 N^T(r)=- N^M(r), \;\;\;m=M_-, & p= (\tau_+ - \tau_-)
             + \int_{-\infty}^{\infty} dr P_M(r) \nonumber \\
T(r)= \tau_+ - \int_{\infty}^{r}dr^\prime P_M(r^\prime )
       & P_T(r)= -M^\prime (r)
\end{eqnarray}
result in (upto total derivatives with respect to $t$) the rexpression of the
parameterized action as
\begin{eqnarray}
S[m, p, M,P_M, R, \Pi_R, N^M, N^R] &
= & \int dt(p{\dot m} + \int_{-\infty}^{\infty}dr(P_T(r){\dot T} (r)
                                + \Pi_R(r){\dot R} (r) )\nonumber\\
& - & \int dt \int_{-\infty}^{\infty}dr (N^T(r)P_T(r) + N^R(r)\Pi_R(r))
\end{eqnarray}
 Thus, $m$ and $p$ are constants of motion and parametrize the reduced
phase space of the theory. $m$ is the ADM mass of the spacetime and
$p$ has the interpretation  of the difference between the proper time,
 $\tau_-$, at left infinity and the parametrization time  at left
infinity with the proper time $\tau_+$ at right infinity synchronized with
the parametrization time at right infinity.

\section{Quantum theory}
 We briefly review Dirac quantization  of the classical
description which follows from the parameterized action (for the quantization
following from the unparameterized action we refer the reader to
\cite{karel}).

Since $R(r),m \geq 0$, it is better to make a point transformation
on the classical phase space before we quantize. We define
\begin{eqnarray}
\xi = \ln R & \Pi_\xi = R\Pi_R \\
x= \ln m  & p_x =mp
\end{eqnarray}
and replace the $\Pi_R=0$ constraint by $\Pi_\xi =0$. To pass to quantum
theory,
we choose a coordinate representation. Thus $\Psi = \Psi (x; T, \xi ]$ ($\Psi$
denotes the wavefunction). The quantum constraints are
\begin{eqnarray}
{\hat P}_T \Psi & = & -i{\delta \Psi \over \delta T(r)} = 0 \\
{\hat \Pi }_\xi \Psi & = & -i{\delta \Psi \over \delta \xi (r)} = 0 \\
\Rightarrow \Psi & = & \Psi (x)
\end{eqnarray}
So the nontrivial operators in the theory are constructed from
${\hat x}$ and ${\hat p}_{x}={\partial \over \partial x}$
and the Hilbert space
consists of  square integrable functions of $x$ on the real line.

\section{Light cone coordinates as canonical variables}
\subsection{Classical theory}
We now pass to a description in terms of canonical variables which correspond
to (on shell) the null `Kruskal' coordinates $U, V$ of
section 2. Recall, from the spacetime solution, that
\begin{equation}
R^2 -{2M\over K} = -K^2 UV \;\;\;\; e^{2KT} = |{V\over U} |
\end{equation}

We start with the description in terms of the parametrized action
$S[m, p, M,P_M, R, \Pi_R, N^M, N^R] $. We define new canonical
variables
\begin{eqnarray}
{\bar m}=m & {\bar R}= R^2 -{2m\over K} & P_{\bar T}= P_T \nonumber \\
{\bar p}= p+\int_{-\infty}^{\infty}dr{\Pi_R \over 2R} &
\Pi_{\bar R} = {\Pi_R \over 2R} & {\bar T} =T
\end{eqnarray}
Motivated by (83) above we make a further canonical transformation (with
${\bar m}, {\bar p}$ unchanged) to new variables ($U,P_U, V,P_V$):
\begin{eqnarray}
{\bar R}= -K^2 UV  &  2K{\bar T}= \ln |V|- \ln |U|  \nonumber \\
P_{\bar R}= -({P_V\over 2K^2 U} + {P_U \over 2K^2 V}) &
P_{\bar T} = KVP_V -KUP_U
\end{eqnarray}

Away from $U=0$ or $V=0$ (which correspond to the horizon on shell),
we can replace the constraints $\Pi_R =0 = P_T$ by
\begin{eqnarray}
P_V & = & 0 \\
P_U & = & 0
\end{eqnarray}
If we impose that $P_U$ and $P_V$
are continuous functions of $r$ on the
constraint surface, they must vanish even on the horizons.
Then the parametrized action becomes
\begin{eqnarray}
S[{\bar m}, {\bar p}, U,P_U, V, P_V, N^U, N^V] &
= & \int dt({\bar p}{\dot{\bar m}} +
\int_{-\infty}^{\infty}dr(P_U(r){\dot U} (r)
                                + P_V(r){\dot V} (r) )\nonumber\\
& - & \int dt \int_{-\infty}^{\infty}dr (N^U(r)P_U(r) + N^V(r)P_V(r))
\end{eqnarray}
where $N^U, N^V$ are the appropriately defined new Lagrange multipliers.
We have refrained from looking at exactly what happens at the horizon and
to what extent it is valid to replace the old constraints with (86) and (87)
because there exists a much better way of defining the null coordinates and
rewriting  the constraints in terms of them \cite{jkm}.
This will be presented elsewhere.

\vspace{2mm}

\noindent Remark:if we  consider the combinations of constraints $H$ and $H_r$
which generate 2 commuting copies of the Virasoro-type algebra, namely
\begin{equation}
H_+ := \Lambda H + H_R \;\;\;\;\; H_-:= \Lambda H - H_r
\end{equation}
then {\em on the constraint surface}, modulo some subtleities on the horizons,
in terms of the new canonical variables
\begin{equation}
H_+ = V^\prime P_V
\end{equation}
\begin{equation}
H_- =-U^\prime  P_U
\end{equation}

\subsection{Quantum theory}
We describe the results of a Dirac quantization of the theory
described by the action above. The fact that $m \geq  0$, is handled by
using ${\bar x} :=\ln {\bar m} $ and $p_{\bar x} := {\bar m}{\bar p}$
(see section 8).
We choose a configuration representation, so that the wave functions
depend on $(U(r), V(r) ,{\bar x})$. The coordinate operators
$ {\hat U},{\hat V},{\hat {\bar x}}$
act by multiplication and the momenta operators
act  as follows:
\begin{eqnarray}
{\hat P}_U & = & -i(\delta / \delta U(r))\nonumber \\
{\hat P}_V & = & -i(\delta / \delta V(r))\nonumber \\
{\hat  p}_{\bar x} & = & -i (\partial / \partial {\bar x})
\end{eqnarray}
The quantum constraints are
\begin{eqnarray}
{\hat P}_U \Psi ({\bar x};U(r),V(r)] & = & 0 \\
{\hat P}_V \Psi ({\bar x};U(r),V(r)] & = & 0 \\
\Rightarrow \Psi = \Psi ({\bar x} )
\end{eqnarray}
The Hilbert space consists of square integrable functions of ${\bar x}$ and
 we have a quantum theory identical to that in section 8.


\section{Conclusions}
In this work, we have formulated the canonical description of vacuum
2-d black holes in terms of variables which have a clear spacetime
interpretation. In doing this we have followed the receipes of \cite{karel}
which dealt with  vacuum Schwarzschild black holes.
As in \cite{karel}, the classical description simplifies to such an
extent that quantization becomes very easy.

When we replace the original classical constraints with new ones, care
has to be taken where the horizon intersects the spatial slice. In particular,
the new constraint functions are assumed to be continuous on the spatial
slice and this forces them to vanish even at the horizon.

 Because the classical variables
used have a clear physical meaning, it becomes easier to interpret the
corresponding quantum operators. We have also constructed, explicitly,
a classical description
(and quantized it) using
the light cone coordinates as canonical
variables. In \cite{karel} the analogs of these objects were the light cone
 Kruskal coordinates and using them explicitly was a little involved
because the Regge -Wheeler tortoise coordinate plays a key role in the
transformation from curvature coordinates to Kruskal coordinates.

We \cite{jkm} are trying to apply methods similar to those used in this
work and in the Schwarzschild case, to the 2d black holes with conformally
coupled matter.

Ultimately, we hope to learn useful lessons from this work and \cite{jkm},
and efforts of other workers like \cite{mikovic,jackiw}
which
will help in tackling the system of  spherically symmetric (4-d) gravity
with a (spherically symmetric) scalar field. \\

\vspace{2mm}

\noindent{\bf Acknowledgements:} I would like to thank Karel Kucha{\v r},
Laszlo Gergely and
Joe Romano for helpful discussions. I would also like to thank Karel
Kucha{\v r} and Joe Romano for
a careful reading of this manuscript. This work was supported by
the NSF grant PHY9207225.

After this work was completed we learnt of a recent
work by Lau\cite{lau} which also contains a  classical analysis
similar to that in this work.

\end{document}